\documentclass[fleqn,12pt,twoside]{article}
\usepackage{espcrc1}

\usepackage{graphicx}

\title{Ultra-Peripheral Collisions in CMS}

\author{Pablo P. Yepes 
\address{Physics and Astronomy Department, MS 315, Rice University \\
  	      Houston, TX 77005, USA},
     for the CMS Collaboration
   	      }

\begin{document}
\maketitle

\begin{abstract}
Coherent peripheral collisions of atomic nuclei involve electromagnetic
or long range hadronic
interactions at impact parameters, where both nuclei survive intact. 
Recently such ultra-peripheral collisions were observed at RHIC.
The effect of the electromagnetic field 
can be interpreted as a photon-photon collision with an effective center of
mass energy up to a few GeV at RHIC. At the Large Hadron Collider
the effective center of mass energy will be
increased by more than an order of magnitude. This opens new   
opportunities, ranging from the study
of non-perturbative QCD to the search for new physics.
\end{abstract}

\date{\today}

\maketitle

\section{Heavy Ions in LHC}

The Large Hadron Collider (LHC) plans to operate as a heavy ion collider for
one month each year starting in 2007. LHC will be capable of accelerating a
variety of ions up to beam energies of 7 TeV/charge. The design, peak 
and three-hour-fill-averaged
luminosities are presented in Table 1 along with the maximum center of mass 
energy\cite{brandt}. 

\begin{table}[hbt]
\begin{tabular}{cccc}
\hline
                         &$L_{max}$&$<$L$>$&$\sqrt{s_{NN}}$ \\
         Ion             &$(cm^{-2}s^{-1})$&$(cm^{-2}s^{-1})$&(GeV)\\
\hline
$^{208}Pb^{82}$          & 1.0 $10^{27}$  & 4.2 $10^{26}$ & 5500 \\
$^{120}Sn^{50}$          & 1.7 $10^{28}$  & 7.6 $10^{27}$ & 5800 \\
$^{84}Kr^{36}$           & 6.6 $10^{28}$  & 3.2 $10^{28}$ & 6000 \\
$^{40}Ar^{18}$           & 1.0 $10^{30}$  & 5.2 $10^{29}$ & 6300 \\
$^{16}O^{8}$             & 3.1 $10^{31}$  & 1.4 $10^{31}$ & 7000 \\
\hline
\end{tabular}
\caption{LHC planned ion beams with design, maximum and 
fill-averaged luminosities along with maximum center of mass energies.}
\label{lhcParameters}
\end{table}

\section{CMS}
The Compact Muon Solenoid (CMS) \cite{cms} experiment is a general-purpose facility to study 
hadronic collisions at the LHC. The detector
consists of a tracking system, electromagnetic and hadronic calorimeters, and muon detectors. A
solenoidal magnet providing a 4 Tesla magnetic field surrounds the tracking and calorimetric
systems. The
tracker, covering the rapidity region $|{\eta}|<$ 2.5, is based on silicon technology. The
electromagnetic calorimeter consists of about 83000 lead-tungstate crystals arranged in 
a central barrel covering
$|{\eta}|<$1.48 and the endcaps, which extend its range to rapidity $|{\eta}|<$3. In the central
barrel, the granularity is as high as ${\eta}$ $x$ $\phi$ = 0.0175 $x$ 0.0175. The hadronic 
calorimeter consists of barrel and endcap sections, each made of sandwiches of copper plates and plastic
scintillator. In the central region ($|{\eta}|<2$) the ${\eta}$ $x$ $\phi$ segmentation is
0.087 $x$ 0.087. The
combination of electromagnetic and hadronic calorimeters provides coverage of the central
rapidity region with excellent energy resolution. In addition, coverage at large rapidity
($3<|\eta|<5$) is achieved by two very forward calorimeters. The large calorimetric coverage 
and good energy resolution makes CMS an optimal detector for jet studies. 
The muon system covers the $|\eta|<2.5$ region. 
In the barrel ($|{\eta}|<1.5$), muons must have a
transverse momentum larger than 3.5 GeV/$c^2$ to be efficiently detected.

\section{Ultra-Peripheral Collisions}
\begin{figure}
\includegraphics[height=.25\textheight]{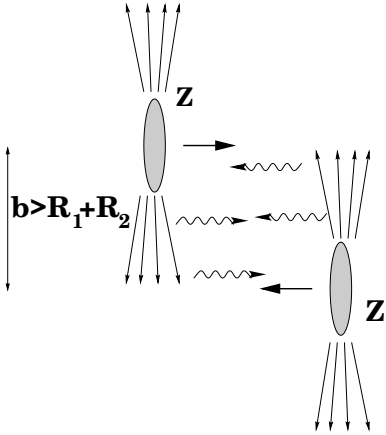}
\caption{Two fast moving nuclei are an abundant source of
quasi-real photons, which can collide with each other and with the
other nucleus. For peripheral collisions with impact parameters $b>R_1+R_2$,
isolated photon-photon and photon-nucleus collisions can be studied.}
\label{twoGammas}
\end{figure}

Nuclear Interactions without direct hadronic collisions are
refered to as Ultra-Peripheral Collisions. The impact parameter 
of the collision
should be larger than the sum of the radii of the two interacting
nuclei. Therefore the only possible 
interactions are due to the electromagnetic 
and/or diffractive processes (See Figure \ref{twoGammas}). For the former
the field is very strong due to the coherent action of all
the protons in the nucleus, 
and the resulting flux of equivalent photons is large, since it is 
proportional to $Z^2$, where the Z is the nuclear charge. 
If the nuclei are required to interact coherently, 
the momentum transfer should be smaller than $\hbar$/L,
where L is the dimension of  the nuclei. In the longitudinal
direction the nuclei are Lorentz contracted by a factor $\gamma$,
and therefore the lontitudinal momentum, $p_L$ must be smaller
than $\gamma~\hbar / R$, where R is the nuclear radius.
For Pb(Ar) at LHC energies $\gamma=2950(3380)$, what translates into a maximum 
longitudinal momentum transfer of about 100(200) GeV. This scale 
determines the maximum mass of the objects produced in this type
of collisions. In the transverse 
plane the nuclei are not contracted, and therefore $p_T<h/R\approx30(60) MeV$ 
for Pb(Ar).
This low value of the total transverse momentum is an important
experimental feature for the separation of signal from background.

\section{Photon-Photon Interactions}
In recent years, considerable progress has been made in understanding
photon-photon collisions in heavy ion collisions. The basic process is 
two virtual (space-like) photons emmitted by the nuclei colliding
to form a final state f. In the equivalent
photon approximaton (EPA), it is assumed that the square of the 4-momenta of
the virtual photons is small ($q_1^2\approx q_2^2 \approx 0$), and therefore
the photons can be treated as quasi-real. This is a good approximation
except for the production of $e^+e^-$ and $\mu^+\mu^-$. When EPA
is utilized the cross section factorizes as the product of the 
$\gamma \gamma \rightarrow f$ cross
section and the $\gamma \gamma$ effective luminosity. 

\begin{figure}
\includegraphics[height=.4\textheight]{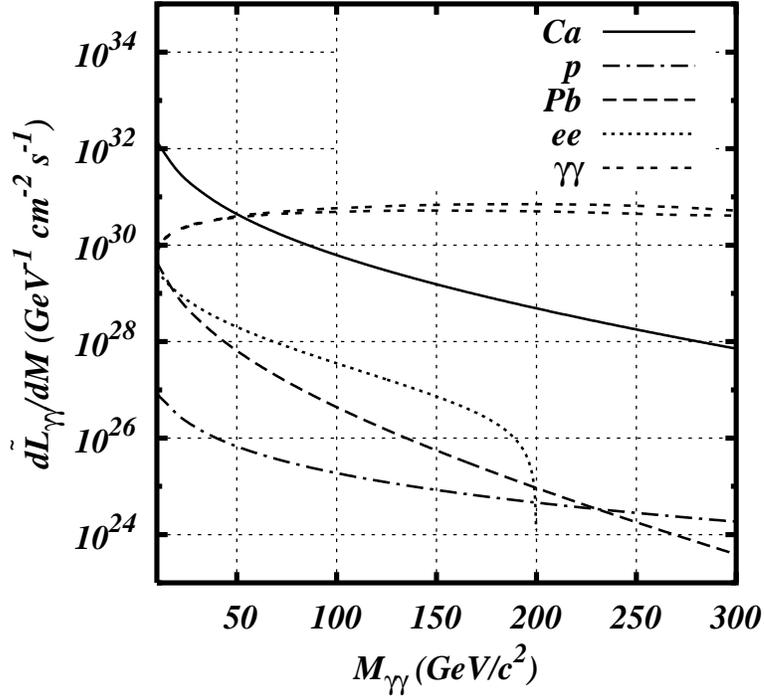}
\label{effectiveL}
\caption{Effective $\gamma\gamma$ luminosities for
Pb+Pb and Ca+Ca at LHC. For comparison the same quantity is shown for
LEP200 and a future Next Linear Collider (NLC),
where photons are obtained by laser backscattering. The following parameters
were used: 
LHC Pb+Pb ($E_{beam}$=2250 GeV, $L=10^{26}cm^{-2}s^{-1}$);
LHC Ca+Ca ($E_{beam}$=3500 GeV, $L=4. 10^{30}cm^{-2}s^{-1}$);
LHC p+p   ($E_{beam}$=7000 GeV, $L=10^{30}cm^{-2}s^{-1}$);
LEP 200 ($E_{beam}$=100 GeV, $L=10^{32} cm^{-2}s^{-1}$);
NLC ($E_{beam}$=500 GeV, $L=2. 10^{33} cm^{-2}s^{-1}$)
}
\end{figure}

Figure \ref{effectiveL} depicts the effective $\gamma\gamma$
luminosities for Pb+Pb and Ca+Ca and LHC as a function of the effective
$\gamma\gamma$ center of mass energy. In addition  
values for proton-proton collisions at LHC, LEP 200 (Large Electron Positron collider) 
and a proposed
Next Linear Collider (NLC) are shown for comparison. As can
be seen, LHC Ca+Ca has the larger effective luminosity for
$M_{\gamma\gamma}<50 GeV$.

\section{Photon-Photon Physics}
Photon-photon interactions open a new window into non-perturbative
QCD, since they provide an independent view of meson and baryon
spectroscopy. They contribute powerful information on the flavor and
spin/angular momentum internal structure of the meson. Using the real
photon approximation, general symmetry principles restric the possible 
final states, in particular spin 1 states are forbiden. In $e^+e^-$ 
annihilation only states with $J^{PC}=1^{--}$ can be produced directly. 
Two photon collisions give access to most of the C=+1 mesons, while C=-1 mesons can
only be produced by the fusion of three or more photons. This allows for
interesting studies of light quark spectroscopy in ultra-peripheral
nuclear collisions, extending the results
already obtained at $e^+e^-$ colliders \cite{berger}. The measurement of meson production 
via $\gamma\gamma$ fusion is also of great interest for glueball searches. The
two-photon width of a resonance is a probe of the charge of its constituents.
Thus the magnitude of the two-photon coupling can serve to distinguish 
quark-dominated from gluon-dominated resonances. Low mass mesons (m<3
2 GeV) are accesible at RHIC energies \cite{kleininpc}. Detailed
studies on the possibilities at LHC have been performed\cite{felixLoI}. 

\begin{figure}
\includegraphics[height=.4\textheight]{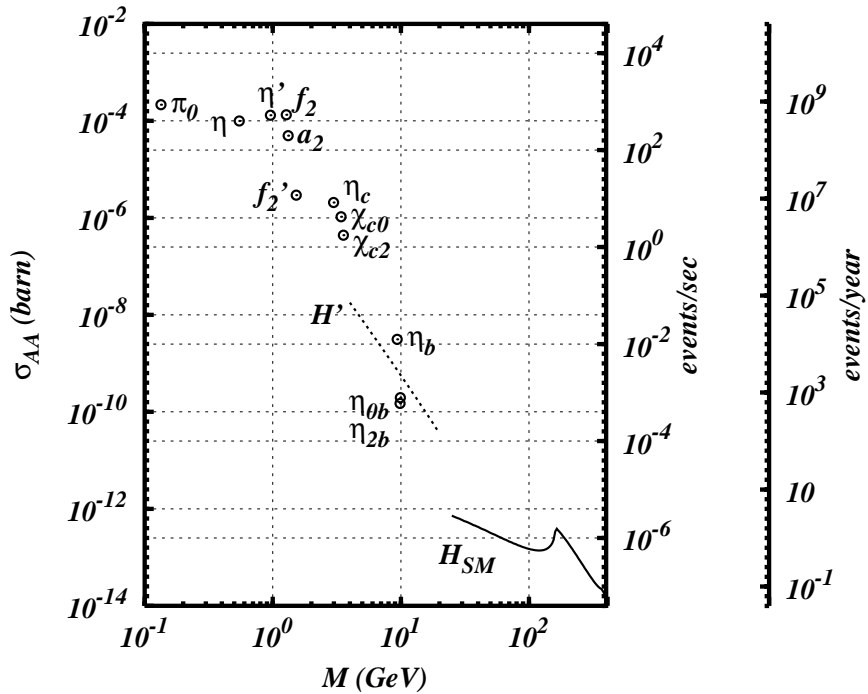}
\caption{Cross sections and production rates, per
second and per year ($10^6$s), of various final states in Ca+Ca collisions
at LHC \cite{cmsHI}.}
\label{resonances}
\end{figure}

Figure \ref{resonances} shows the production cross section, and rates per
second and year, for differente final states in Ca+Ca interactions at LHC.
As can be seen, millions of C-even charmonium states will be produced in
a one-month ($10^6$ s) run. Assuming a detector 
efficiency of 10\%, the number of expected charmonium events in Ca+Ca(Pb+Pb)
collisions is $\approx 10^6$($5. 10^3$). This is three orders of magnitude larger
than what was expected during five years of running at LEP2000.

\begin{figure}
\includegraphics[height=.4\textheight]{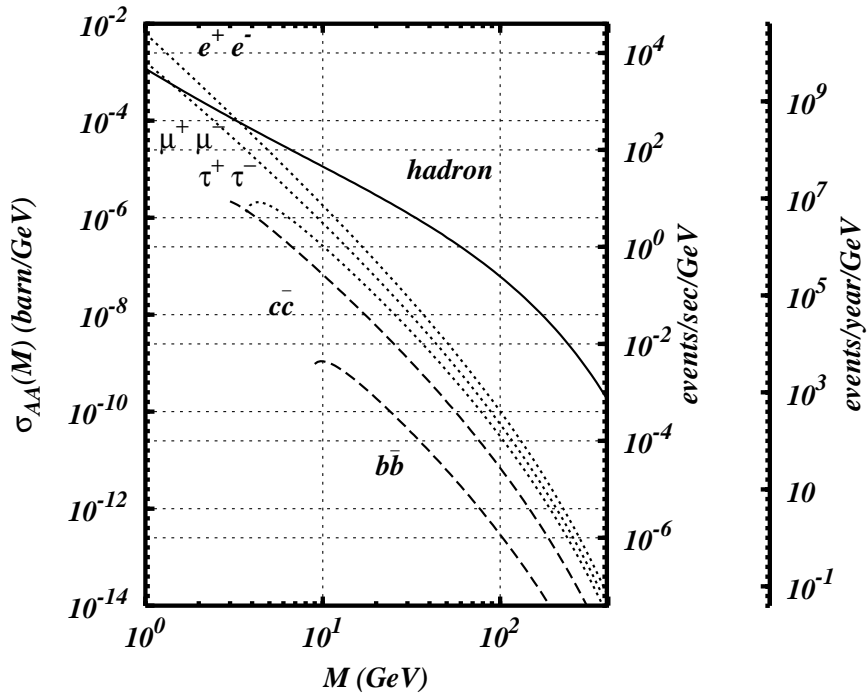}
\caption{Cross sections and production rates, per second and per year
($10^6$s), per GeV for different dileptons and $Q\bar{Q}$ pairs in
Ca+Ca collisions at LHC. Also shown is the total hadronic cross section.
\cite{cmsHI}.
}
\label{continuum}
\end{figure}

There are various mechanisms to produce hadrons in photon-photon collisions.
It is of great theoretical interest to understand the relative importance of
those mechanisms and their properties. LEP has measured the total 
$\gamma\gamma\rightarrow hadrons$ cross section to invariant masses up to 70 GeV. 
Figure \ref{continuum} depicts cross sections for hadronic production
along with di-lepton and $Q\bar{Q}$ production. As can be seen, the production
rates per year are large, what will allow CMS to make important 
contributions in this field.

The high effective $\gamma\gamma$ luminosities at LHC also offers interesting
possibilities for the search of new physics. It has been proposed to look
for the Standard Model Higgs in these collisions. Unfortunately the rates
are of only a few events a year, as can be seen on Figure \ref{resonances}.
However alternative scenarios with light Higgs bosons with larger couplings
to $\gamma\gamma$ are possible and have been proposed in the literature
\cite{higgsModels}. Any new particles with strong couplings to the
$\gamma\gamma$ channel, will have large production cross sections
\cite{gammaWidth,ggW2}. Since the $\gamma\gamma$ width of a resonance is
mainly proportional to the wave function at the origin, huge values can be
obtained for very tightly bound systems, like composite scalar bosons.
Therefore the search for this kind of resonances in the $\gamma\gamma$ channel
will be possible at the LHC.

\section{Photon-Nucleus}
The ep collider HERA has studied $\gamma p$ interactions up to center of masses 
$W_{\gamma p}$=200 GeV in great detail. At LHC photon-nucleus
interactions, with a center of mass energy up to $W_{\gamma p}$=950 GeV, will
be produced with sizeable cross sections \cite{baur}. Moreover 
$\gamma A$ interaction rates are expected to be very large
at LHC. As an example, Table \ref{vectorMeson}
shows the rates for coherent $\gamma A \rightarrow f$, with $f=\rho^0, \omega, \phi$
and J/$\Psi$ for AuAu collisions at RHIC, along with Pb+Pb and Ca+Ca at LHC
\cite{kleinVectorMeson}. 
These very large rates turn relativistic heavy
ion colliders into vector meson factories, with rates that may be competitive
with accelerators dedicated to this physics. 
Recently the STAR collaboration has reported the
observation of the $\rho$ meson in coherent $\gamma A$ collisions \cite{kleininpc}, 
showing that it is posible to separate those events from the background from 
peripheral hadronic events. 
A more comprehensive discussion on 
the $\gamma A$ physics at CMS can be found elsewhere \cite{cmsHI}.
\begin{table}[hbt]
\begin{tabular}{cccc}
\hline
     Meson               & RHIC-Au & LHC-Pb & LHC-Ca \\
\hline
$\rho^0$                 & 120   & 520 & 230000 \\
$\omega$                 &  12   &  49 &  23000 \\
$\phi  $                 & 7.9   &  46 &  15000 \\
J/$\Psi$                 & 0.058 & 3.2 &    780 \\
\hline
\end{tabular}
\caption{Meson production rates, in Hz, at design luminosity
for various beams \cite{kleinVectorMeson}.}
\label{vectorMeson}
\end{table}
\section{Conclusions}
LHC plans to provide one month of heavy ion running every year. This
opens unique opportunities for the study of photon-photon and photon-nucleus 
collisions. CMS with with its large acceptance is well equipped to address many
of the physics topics of this field.
\bibliography{sample}

\end{document}